\documentclass[twocolumn,showpacs,amsmath,nofootinbib,pra,aps,amssymb,longbibliography]{revtex4-1}

\usepackage[T1]{fontenc}
\usepackage[utf8]{inputenc}
\usepackage{times}
\usepackage{color} 
\usepackage{array}
\usepackage{amssymb,amsmath}
\usepackage{amsbsy}
\usepackage[pdftex]{graphicx} 
\usepackage{bm}
\usepackage{float}
\usepackage{dcolumn}
 
\usepackage[unicode,breaklinks]{hyperref}
\hypersetup{
    unicode=true,
    plainpages=false, 
    colorlinks=true,
    linkcolor=blue,
    citecolor=blue,
    filecolor=black,
    urlcolor=blue
}
\usepackage{url}
\usepackage{verbatim}

\begin{document}
 
\title{Superfluid fraction tensor of a two-dimensional supersolid} 
	\author{P.~B.~Blakie  }
	
	\affiliation{%
	 Department of Physics, University of Otago, Dunedin 9016, New Zealand}
\date{\today} 
\begin{abstract}   
We investigate the superfluid fraction of crystalline stationary states within the framework of mean-field Gross-Pitaevskii theory. Our primary focus is on a two-dimensional system with a non-local soft-core interaction, where the superfluid fraction is described by a rank-2 tensor. We analyze and establish connections between methods for calculating the superfluid tensor derived from analysis of the nonclassical translational inertia and the effective mass. We then apply these methods for crystalline states exhibiting triangular, square, and stripe geometries across a broad range of interaction parameters. Factors leading to an anisotropic superfluid fraction tensor are also considered. We also refine the Leggett bounds for the superfluid fraction to an accurate approach that involves a calculation using the density profile over a single unit cell. We systematically compare these  bounds to our full numerical results, and other results in the literature. This work is of direct relevance to other supersolid systems of current interest, such as supersolids produced using dipolar Bose-Einstein condensates.
\end{abstract} 

\maketitle
\section{Introduction}
Superfluidity emerges as a distinctive state of matter within quantum many-body systems, manifesting under specific conditions and exhibits various traits, most notably the absence of viscosity. An enduring question has been whether superfluidity could coexist with crystalline order realizing a supersolid state of matter \cite{Andreev1969a,Chester1970a,Leggett1970a,Boninsegni2012a}. In recent years several different types of experiment utilizing Bose-Einstein condensates (BECs) have successfully observed supersolidity 
\cite{Leonard2017a,Li2017a,Tanzi2019a,Bottcher2019a,Chomaz2019a}.

 Leggett's seminal work highlighted that the introduction of crystalline order, with the concomitant breakdown of translational invariance, leads to a reduction in the system's superfluid fraction even at zero temperature \cite{Leggett1970a,Leggett1998a}. Although the  emergence of crystalline order has been directly witnessed in supersolids produced using BECs, the characterization of superfluid transport remains less straightforward. For dipolar BEC supersolids, excitation spectroscopy has been employed to infer superfluidity \cite{Natale2019a,Guo2019a,Tanzi2019b} (also see \cite{Saccani2012a,Macri2013a}). However, this does not directly quantify the superfluid fraction of the system. We also note that scissor mode excitations have furnished evidence of the diminished superfluid fraction  \cite{Tanzi2021a}.

Recently two groups have quantified the superfluid fraction in  an artificial supersolid  produced by applying an optical lattice to a BEC \cite{Tao2023a,Chauveau2023a} (i.e.~with imposed, rather than spontaneously formed density order).  Their superfluid measurements were based on two methods: (i) a superfluid bound developed by Leggett, which requires the precise measurement of the density profile \cite{Chauveau2023a};  and (2) a ratio of the speeds of sound propagating parallel and perpendicular to the lattice used to obtain the effective mass (and thereby the superfluid fraction) along the lattice \cite{Tao2023a,Chauveau2023a}.    
It is currently not clear how to extend these ideas  to two-dimensional (2D) or three-dimensional (3D) supersolids, where the superfluid fraction is properly a rank-2 tensor.  For example, the sound measurements do not generally apply to supersolids where spontaneous translational symmetry breaking leads to multiple gapless sound branches \cite{Watanabe2012a,Roccuzzo2019a,Hertkorn2019a,Blakie2023a}.   The first experimental observation of 2D dipolar supersolids \cite{Norcia2021a} underscores the need for a deeper understanding of superfluidity in higher dimensions.

In this work we consider theoretical methods to quantify the superfluid fraction for a 2D supersolid. We use meanfield model of a 2D Bose gas with soft-core interactions to illustrate our study. This model exhibits a transition to a triangular crystal when the dimensionless interaction parameter, $\Lambda$, surpasses a threshold value \cite{Pomeau1994a}. Furthermore, our investigation encompasses metastable crystal configurations characterized by square or stripe geometries, thereby furnishing a diverse platform to implement and validate superfluid measures.
The three main components of this work are: 
(1) We review the definition of the superfluid tensor via the nonclassical translational inertia. This describes the steady state response of the system to moving walls, which carries the normal fluid but leaves the superfluid at rest. For the crystalline ground state the superfluidity can be quantified by understanding the phase profile in a unit cell in response to the motion. The equations determining this phase profile were presented in Ref.~\cite{Josserand2007a}, but are equivalent to a formulation by Saslow  in 1976 \cite{Saslow1976a}.  The superfluid tensor for the 2D soft-core system was previously computed using this approach in Ref.~\cite{Sepulveda2010a}, however our results disagree with these, underscoring the complexity of the calculation.
(2) We examine the effective mass of the BEC in a crystalline state, which also relates to the superfluid fraction \cite{Pitaevskii2004a}. This approach has been used to determine the superfluid fraction tensor in a 2D case by solving the nonlinear Gross-Pitaevskii equation (GPE) for the condensate in motion \cite{Hsueh2017a}. We discuss the connection to the phase profile formulation and show that the effective mass can be determined by a simpler linearized problem, which is equivalent to solving for the Hartree excitations. (3)  The Leggett upper bound was used by Zhang \textit{et al.}~\cite{Zhang2019a} to quantify the superfluid fraction of a 2D dipolar supersolid, although this involved minimising over angles, and no comparison was made to a direct calculation of the superfluid fraction to quantify the reliability of the bound. Here we comprehensively study the use of the Leggett upper and lower bounds in application to a 2D supersolid. From this formulate a useful form of the bound involving a single unit cell without the need to consider various angles. Also, our general techniques allow us to explore conditions under which 2D crystalline states exhibit anisotropic superfluidity  (cf.~\cite{Hsueh2017a}), hence revealing situations where the tensorial nature of the superfluid fraction manifests.

\section{2D Soft-core supersolid model}\label{Sec:2Dmodel}
Our model is a meanfield description of a 2D BEC, without any confining trap, and interacting via a soft-core interaction potential  $U(\mathbf{x})=U_0\theta(a_\mathrm{sc}-|\mathbf{x}|)$, where $a_\mathrm{sc}$ is the core radius, $U_0$ is the potential strength, $\theta$ is the Heaviside step function, and $\mathbf{x}=(x,y)$.   The meanfield energy functional for this system is
\begin{align}
E\!=\!\int\!d^2\mathbf{x}\,\Psi^*\!\left(\!-\frac{\hbar^2}{2m}\nabla^2\!+\!\frac{1}{2}\!\int d^2\mathbf{x}^\prime\,U(\mathbf{x}\!-\!\mathbf{x}^\prime)|\Psi(\mathbf{x}^\prime)|^2\right)\Psi,\label{fullE}
\end{align}
where the wavefunction $\Psi$ is normalized to the total atom number $N$.
Here we will be interested in the ground state for $N$ atoms in a large plane of area $A$, in the limit that both of these are large, and with a mean areal density of $n=N/A$. We then choose to set $\Psi(\mathbf{x})=\sqrt{n}\psi(\mathbf{x})$, so that  
$\int  d^2\mathbf{x}\,|\psi(\mathbf{x})|^2=A.$

The interaction term is efficiently evaluated by the convolution theorem, where the $k$-space form of the interaction potential is $\tilde{U}(\mathbf{k})=2\pi U_0a_\mathrm{sc}J_1({k}_\rho a_\mathrm{sc})/{k}$.  
Adopting $a_\mathrm{sc}$ as the unit of length and $\hbar\omega_0=\hbar^2/ma_{\mathrm{sc}}^2$ as the unit of energy, it is convenient to define the dimensionless interaction  parameter
\begin{align}
\Lambda =\frac{\pi na_{\mathrm{sc}}^2U_0}{\hbar\omega_0}.
\end{align}

We look for spatially periodic (crystalline) stationary state solutions. We consider 2D periodic solutions on a unit cell with lattice vectors $\mathbf{a}_1=a \hat{\mathbf{x}}$, and $\mathbf{a}_2=a(\cos\theta \hat{\mathbf{x}}+\sin \theta  \hat{\mathbf{y}})$, where we will consider $\theta=\pi/3$ (triangular lattice) or $\theta=\pi/2$ (square lattice). We also consider a 1D stripe phase with a single lattice vector $\mathbf{a}_1=a \hat{\mathbf{x}}$, with $\Psi(\mathbf{x})=\sqrt{n}\psi(x)$. In both the 1D and 2D cases, the stationary states are characterized by  a single cell size parameter $a$.

Numerically the stationary states can be calculated with spectral accuracy using a planewave presentation in reciprocal space associated with the unit cell we choose to consider (e.g.~see \cite{Kunimi2012a}). We use the gradient flow method \cite{Bao2004a,Lee2021a} to minimise the energy and obtain the ground state solution for the unit cell. We note that these solutions satisfy the time-independent GPE 
$\mu\psi=\mathcal{L}_{\mathrm{GP}}\psi$, where
\begin{align} 
\mathcal{L}_{\mathrm{GP}}=-\frac{\hbar^2}{2m}\nabla^2+n\Phi(\mathbf{x}),\label{LGP}
\end{align}
is the GP operator and $\mu$ is the chemical potential.

We denote the ground state energy per particle for a cell of size $a$ as 
\begin{align}
\mathcal{E}(n;a)=\frac{1}{A_{\mathrm{uc}}}\int_\mathrm{uc} d^2\mathbf{x}\,\psi^*\left(-\frac{\hbar^2}{2m}\nabla^2+\frac{n}{2}\Phi(\mathbf{x})\right)\psi,\label{Ena}
\end{align}
where uc denotes the unit cell, $A_{\mathrm{uc}}=|\mathbf{a}_1\times\mathbf{a}_2|$ is the unit cell area and
\begin{align}
\Phi(\mathbf{x})=\int d^2\mathbf{x}^\prime\,U(\mathbf{x}-\mathbf{x}^\prime)|\psi(\mathbf{x}^\prime)|^2.\label{Eq:Phi}
\end{align}
For the uniform state the energy per particle is independent of the unit cell and is given by $\mathcal{E}_u=\frac{1}{2}\Lambda \hbar\omega_0$.  
For the modulated states there is an optimal value of cell size
$\mathcal{E}(n)\equiv\min_a \mathcal{E}(n;a)$, that minimises the energy per particle.

\begin{figure}[htbp]
	\centering
	\includegraphics[width=3.35in]{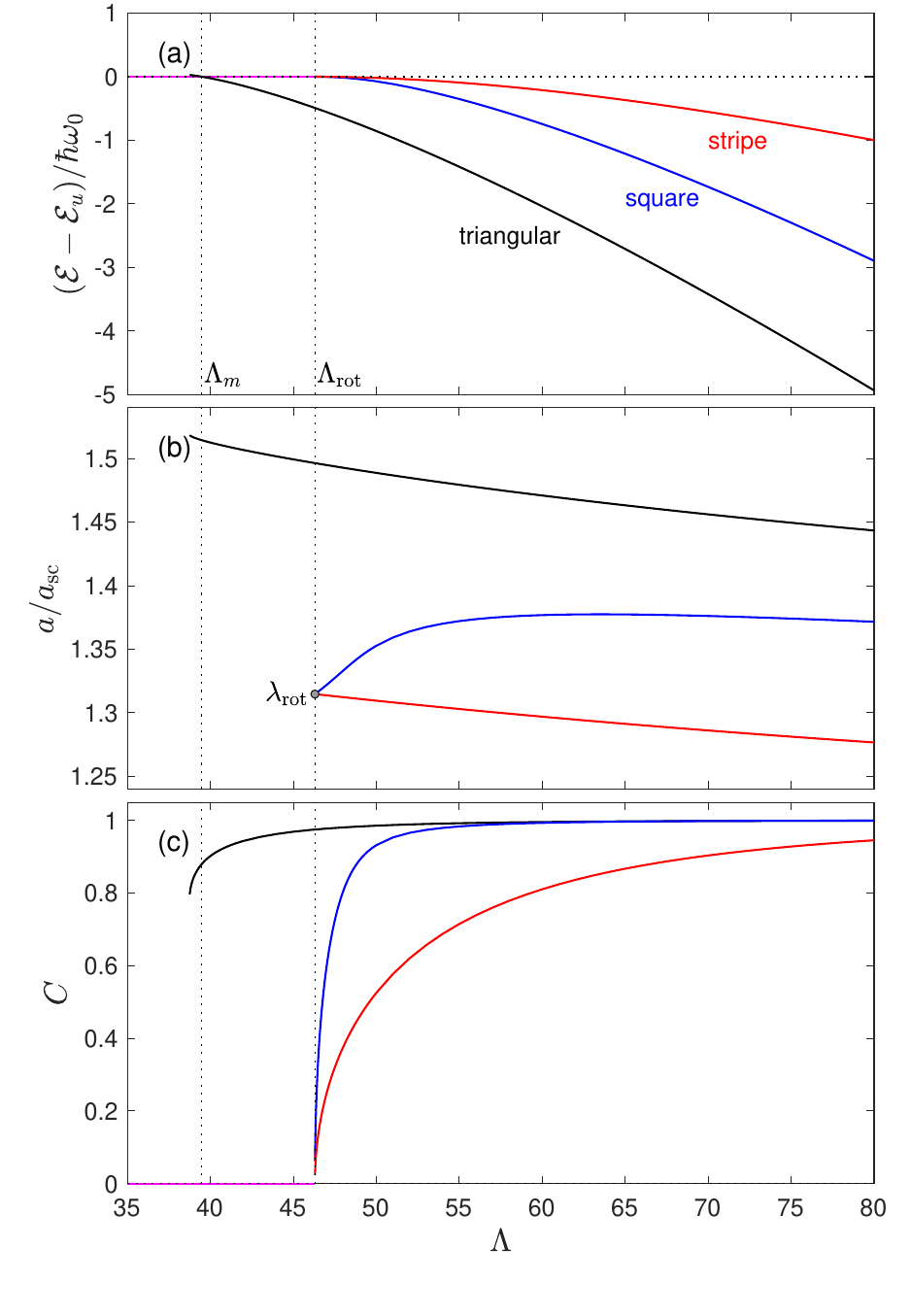}
	\caption{Crystallisation phase transition of a 2D soft-core BEC. (a) The energy per particle relative to the uniform ground state for the triangular (black), square (blue), stripe (red) and uniform (magenta) states. (b) The energy minimising unit cell size of the modulated states. (c) The density contrast. 
	Vertical dotted lines indicate the melting point for the triangular state  and the critical point for the square and stripe states, which coincides with the roton softening of the uniform state. In (b) the roton wavelength is indicated for reference.
	\label{figCfs}}
\end{figure}

The phase diagram of this model is well-known (see \cite{Pomeau1994a,Sepulveda2010a,Watanabe2012a,Kunimi2012a,Macri2013a,Prestipino2018a}) and is summarised in Fig.~\ref{figCfs}. As $\Lambda$ changes the ground state undergoes a first order phase transition from the uniform state to a triangular state at $\Lambda_m=39.49$ [see Fig.~\ref{figCfs}(a)]. The stripe and square states are metastable states, and undergo a second order transition to the uniform state at $\Lambda_{\mathrm{rot}}=46.298$, where a roton excitation softens in the uniform state. 
The optimal unit cell size $a$ for the modulated states is shown in Fig.~\ref{figCfs}(b). At the transition point the cell size of  the square and stripe states matches the roton wavelength. In  Fig.~\ref{figCfs}(c) we show the density contrast 
\begin{align}
C=\frac{|\psi|^2_{\max}-|\psi|^2_{\min}}{|\psi|^2_{\max}+|\psi|^2_{\min}},
\end{align}
 with $|\psi|^2_{\max}$ and $|\psi|^2_{\min}$ being the maximum and minimum values of $|\psi|^2$ in the unit cell, respectively. The density contrast is 0 in the uniform state, and thus serves as an order parameter for crystalline order.

\section{Superfluidity}

\begin{figure}[htbp]
	\centering
	\includegraphics[width=3.2in]{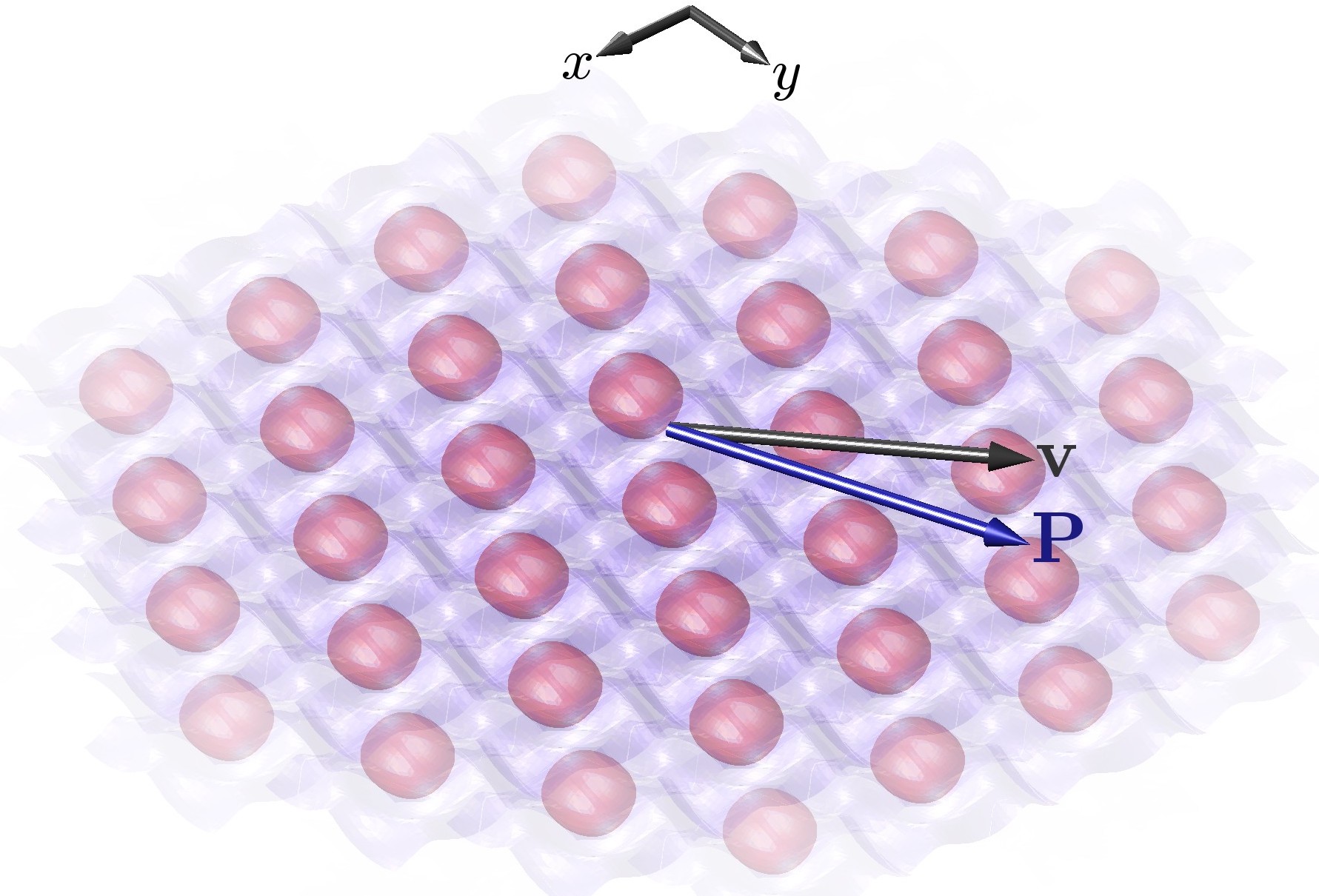}
	\caption{Schematic of moving frame used to  define superfluidity. The walls are taken to move at a velocity of $\mathbf{v}$ relative to the lab frame, and in steady-state the crystalline structure will move with the walls.  The superfluid remains stationary in the lab frame. The momentum $\mathbf{P}$ of the moving system can be in a different direction to $\mathbf{v}$ due to the tensorial character of the superfluidity. 
	\label{figmoveframe}}
\end{figure}

Having introduced our model we now turn to consider how to evaluate the superfluid fraction. We focus on the 2D case, relevant to our model, although our results  generalize to 1D and 3D cases.

\subsection{Supersolid in a slowly moving frame}\label{Sec:mvframe}
To understand the superfluid response we consider our system with fictitious walls moving at a velocity of $\mathbf{v}$ with respect to the lab frame (see Fig.~\ref{figmoveframe}). The moving walls define the preferred frame for the normal component. We compute the ground state in the co-moving frame by solving the time-independent GPE
\begin{align}
\mu_\mathbf{v}\psi_\mathbf{v}= 
\mathcal{L}_{\mathrm{GP}}\psi_\mathbf{v}
+i\hbar\mathbf{v}\cdot\bm{\nabla}\psi_\mathbf{v}.
\end{align}
For small $v$ we can make a perturbative expansion for the spatially periodic solution  
\begin{align} 
\psi_\mathbf{v}(\mathbf{x})\approx\psi_0(\mathbf{x})e^{i\phi_1(\mathbf{x})},\label{psivphi1}
\end{align}
 where $\psi_0$  is the  stationary solution for $\mathbf{v}=0$ and $\phi_1$ is the first order correction to the phase.  
  We emphasize that changes in the density profile occur  beyond first order and are neglected in our treatment. 
In the moving frame the current is 
\begin{align}
\mathbf{j}=\rho_0(\mathbf{x})\left(\frac{\hbar}{m}\bm{\nabla}\phi_1-\mathbf{v}\right),
\end{align} 
where   $\rho_0\equiv n|\psi_0|^2$. 
The  phase term is determined from the condition $\bm{\nabla}\cdot\mathbf{j}=0$, which ensures the density is stationary in the moving frame (i.e.~the crystal is co-moving). In 1D this condition means that the current is constant. In higher spatial dimensions non-trivial flow patterns can emerge,  determined by solving
\begin{align}
\frac{\hbar}{m}\bm{\nabla}\cdot(\rho_0 \bm{\nabla}\phi_{1})=\mathbf{v}\cdot\bm{\nabla}\rho_0.\label{linphi1}
\end{align}   
 Saslow \cite{Saslow1976a} originally obtained this result, albeit in reciprocal space, and used it to estimate the superfluid fraction of solid $^4$He (also see \cite{Saslow2012a}).
 
From Eq.~(\ref{psivphi1}) the momentum in the lab frame is  
 \begin{align}
 \mathbf{P}(\mathbf{v})= \hbar\int d^2\mathbf{x}\,\rho_0 \bm{\nabla}{\phi_1},\label{Plab}\end{align}
 which can potentially be in a different direction to $\mathbf{v}$ (see Fig.~\ref{figmoveframe}). The energy of this state is 
\begin{align}
E(\mathbf{v})&=\frac{\hbar^2n}{2m}\int  d^2\mathbf{x}\,\bm{\nabla}\psi_\mathbf{v}^*\cdot \bm{\nabla}\psi_\mathbf{v}+E_{\mathrm{int}}(\rho_0),\\
&=E_0+\frac{\hbar^{2}}{2m}\int  d^2\mathbf{x}\,\rho_0 |\bm{\nabla}\phi_1|^2,\label{Elab}
\end{align}
where $E_{\mathrm{int}}(\rho_0)=\frac{1}{2}\int d^2\mathbf{x}\,n\Phi(\mathbf{x})\rho_0(\mathbf{x})$ denotes the interaction energy and $E_0$ is the energy\footnote{The kinetic energy of $\psi_\mathbf{v}$ is $\frac{\hbar^{2}}{2m}\int d^2\mathbf{x}\,\left[\frac{1}{4}\frac{(\bm{\nabla}\rho_0)^{2}}{\rho_0}+\rho_0(\bm{\nabla}\phi_1)^{2}\right]$. The first term is the energy of $\psi_0$, the next term  is the additional energy of $\psi_\mathbf{v}$.} of  $\psi_0$.  

If the integration region is taken over a whole number of unit cells, then using integration by parts and Eq.~(\ref{linphi1}), we obtain $\int  d^2\mathbf{x}\,\rho_0 |\bm{\nabla}\phi_1|^2=\frac{m}{\hbar}\mathbf{v}\cdot\int d^2\mathbf{x}\,\rho_0 \bm{\nabla}{\phi_1}$. Under these conditions Eq.~(\ref{Elab}) can be written as
\begin{align}
E(\mathbf{v})&=E_0+\tfrac{1}{2}\mathbf{v}\cdot \mathbf{P}(\mathbf{v}).\label{ElabP}
\end{align}

\subsubsection{Linear solution}
The linear system (\ref{linphi1})  can be recast by introducing the auxiliary vector function $\mathbf{K}=(K_x,K_y)$ as  
\begin{align}
\phi_{1}=\frac{m}{\hbar}\mathbf{v}\cdot \mathbf{K}(\mathbf{x}).\label{phi1K}
\end{align}
 The component functions satisfy [from Eq.~(\ref{linphi1})]
\begin{align}
 \bm{\nabla}\!\cdot\left(\rho_0 \bm{\nabla}K_i\right)&=\frac{\partial\rho_0 }{\partial x_i},\label{Ki}
\end{align}
where we use the index notation $i=\{x,y\}$. In terms of the auxiliary vector function, the lab frame momentum (\ref{Plab}) is
 \begin{align}
 \mathbf{P}(\mathbf{v})= m\int d^2\mathbf{x}\,\rho_0 \bm{\nabla}(\mathbf{v}\cdot\mathbf{K}). \label{PlabK}
 \end{align} 
 
For a given density $\rho_0$, we can solve Eq.~(\ref{Ki}) for $\mathbf{K}$ by inverting this linear system\footnote{Eq.~(\ref{Ki}) is singular: a constant function is a non-trivial solution of the homogeneous problem (reflecting invariance of $\psi$ to a global phase change). This can be remedied by transforming (\ref{Ki}) to reciprocal space and projecting out the $\mathbf{k}=\mathbf{0}$ component.}. Because  $\rho_0$ and $\mathbf{K}$ are periodic we can solve Eq.~(\ref{Ki}) using a single unit cell.
Equation (\ref{Ki})  was obtained by Josserand \textit{et al.~}\cite{Josserand2007a,Josserand2007b,Sepulveda2008a,Sepulveda2010a} by applying the technique of homogenization to the GPE in the context of studying the superfluidity in soft-core models\footnote{Our definition of $\mathbf{K}$ differs in sign from \cite{Sepulveda2008a,Sepulveda2010a}.}.

\subsubsection{Energy response to moving walls}\label{Sec:ESF}
The superfluidity can be identified from the lab frame energy.  Writing this energy the  form \cite{Leggett1970a}
\begin{align}
E(\mathbf{v})=E_0+\frac{1}{2}Mv^2+\Delta E(\mathbf{v}),\label{Emovingwall}
\end{align}
where $M=Nm$ is the classical inertia of the system and $N$ is the atom number. The term $\Delta E(\mathbf{v})$ describes the departure of the energy from the classical result $(\frac{1}{2}Mv^2)$, associated with the superfluid component. From this we can define the superfluid fraction tensor $f_{ij}$ as 
\begin{align}
\lim_{v\to0}\Delta E(\mathbf{v})=-\frac{1}{2}M\sum_{ij}f_{ij}v_iv_j,\qquad   i,j=\{x,y\} .\label{ESF}
\end{align}
Notably, when $f_{ij}\to\delta_{ij}$ the system is a pure superfluid and $E(\mathbf{v})$ is independent of $\mathbf{v}$ for sufficiently small velocities. 

Comparing Eqs.~(\ref{ElabP}) and (\ref{Emovingwall}), we see that 
\begin{align}
\Delta E(\mathbf{v})=\frac{1}{2}\mathbf{v}\cdot\left[\mathbf{P}(\mathbf{v})-M\mathbf{v}\right],\label{DEP}
\end{align}
 and using (\ref{PlabK}) we obtain
\begin{align} 
 f_{ij}=-\frac{1}{M}\frac{\partial^2 \Delta  E}{\partial v_i\partial v_j}=\delta_{ij}-\frac{1}{N}\int\! d^2\mathbf{x}\,\rho_0\! \left(\frac{\partial K_j}{\partial x_i}\right).\label{fijE} 
\end{align} 
 
\subsubsection{Momentum response to moving walls}\label{Sec:PSF}
We can also consider the momentum of the system in the lab frame, associating this momentum with the normal component that moves with the walls. In the low velocity limit this leads to the identification
\begin{align}
\lim_{v\to0}P_i(\mathbf{v})=M\sum_{j}(\delta_{ij}-f_{ij})v_j,\label{PSF}
\end{align}
such that for $f_{ij}=0$ we have $\mathbf{P} =M\mathbf{v}$. 
Thus the superfluid fraction, using Eq.~(\ref{PlabK}),   is
\begin{align}
 f_{ij}&=\delta_{ij}-\frac{1}{M}\frac{\partial P_i}{\partial v_j}=\delta_{ij}-\frac{1}{N }\int  d^2\mathbf{x}\,\rho_0 \frac{\partial K_{j}}{\partial x_{i}},\label{fijP}
\end{align} 
which is identical to result (\ref{fijE}).

\subsection{Superfluid phase twist and effective mass}
We now discuss an alternative approach to identifying the superfluid tensor.
To do this we consider the system in the frame where the crystal is stationary, and apply a phase twist to the wavefunction. 
For our 2D system we can specify the direction of phase twist as 
$\Delta \bm\theta=\Delta\theta_x\hat{\mathbf{x}}+\Delta\theta_y\hat{\mathbf{y}}$, over some macroscopic sample length of $L$. We assume this twist manifests as an average phase gradient that can be characterized by the quasimomentum $\mathbf{q}=\Delta\bm{\theta}/L$. 
For small $q$ the energy, which we denote at $E^\prime(\mathbf{q})$, is (to second order in $\mathbf{q}$) 
\begin{align}
E^\prime(\mathbf{q})=E_0+ N\sum_{ij}\frac{{\hbar^2}q_iq_j}{2m^*_{ij}},
\end{align}
which defines the effective mass tensor ${m^*_{ij}}$. 
 The relationship between the superfluid and effective mass tensors is
\begin{align}
f_{ij}=\frac{m}{m^*_{ij}}=\frac{m}{N\hbar^2}\left(\frac{\partial ^2 E^\prime}{\partial q_i \partial q_j}\right),\label{fijmstar}
\end{align}
as discussed in \cite{Eggington1977a,Pitaevskii2004a,Hsueh2017a,Chauveau2023a}.
Here, motivated by this definition, we explore an alternative method to compute the superfluid fraction and then discuss the equivalence of (\ref{fijmstar}) to our earlier formulation.

\subsubsection{Calculation of effective mass of a supersolid} 
The ground state with quasimomentum $\mathbf{q}$ is a Bloch wave solution of the GPE 
\begin{align}\mu_\mathbf{q}\psi_\mathbf{q}= 
\mathcal{L}_{\mathrm{GP}}\psi_\mathbf{q}.
\end{align} 
Similar to the analysis of Sec.~\ref{Sec:ESF}, we find that to leading order in $\mathbf{q}$ the density of the ground state is unchanged, so in this limit we can set 
\begin{align}
\psi_{\mathbf{q}}(\mathbf{x})=\psi_{0}(\mathbf{x})e^{i\mathbf{q}\cdot{\mathbf{x}}+i\phi_1(\mathbf{x})}.\label{psiq}
\end{align}
This solution obeys Bloch's theorem where $\psi_{0}(\mathbf{x})e^{i\phi_1(\mathbf{x})}$ is periodic on the unit cell. Thus we see that $\phi_1(\mathbf{x})$ describes the response to the quasimomentum phase gradient arising from the crystalline density. In 1D this can lead to the overall phase profile having a staircase pattern  (see Ref.~\cite{Tao2023a}).

Also, the unchanged density profile means that we can set $\Phi\to
\Phi_0\equiv\int d^2\mathbf{x}^\prime\,U(\mathbf{x}-\mathbf{x}^\prime)|\psi_0(\mathbf{x}^\prime)|^2$. Thus $\psi_{\mathbf{q}}$ can be determined by solving the linearized GP equation \begin{align}
\mu_{\mathbf{q}}\psi_{\mathbf{q}}=\left(-\frac{\hbar^2}{2m}\nabla^2+n\Phi_0(\mathbf{x})\right)\psi_{\mathbf{q}},\label{EGPq}
\end{align}  
which coincides with the single particle Hartree excitations for the lowest band.
Comparing the full energy functional (\ref{fullE}) to the GPE operator (\ref{LGP}) we see the energy these Bloch states can be evaluated using the result 
\begin{align}E^\prime(\mathbf{q})=N\mu_\mathbf{q}-\frac{1}{2}E_{\mathrm{int},0},
\end{align}
with $E_{\mathrm{int},0}$ being the interaction energy calculated with $\psi_0$ (noting that $E_{\mathrm{int},0}$ is independent of $\mathbf{q}$). 
Since methods for solving the Schr\"odinger equation are well developed, it is relatively straightforward to numerically evaluate $\mu_\mathbf{q}$ and hence determine the effective mass by computing finite difference second derivatives of $N\mu_{\mathbf{q}}$ with small changes in $\mathbf{q}$.

\subsubsection{Relation to moving frame treatment}
It is worth considering the relationship to the analysis presented in Sec.~\ref{Sec:mvframe}. In defining the effective mass we are working in the (co-moving) frame where the crystal is stationary, and the superfluid moves with an average superfluid velocity of $\mathbf{v}_s=\hbar \mathbf{q}/m$. Equivalently, the crystal is moving at a velocity of $\mathbf{v}=-\hbar \mathbf{q}/m$ relative to the superfluid (lab) frame.  First, this observation reveals that $\psi_{\mathbf{q}}$, as given in (\ref{psiq}), in just result (\ref{psivphi1}) boosted by $\mathbf{v}$ to the crystal frame. This also confirms that the function $\phi_1$ appearing in both treatments is identical.

The previous result we obtained for the lab frame energy  $E(\mathbf{v})$  [Eq.~(\ref{ElabP})] and $E^\prime(\mathbf{q})$, are thus related by the Galilean transformation  
\begin{align}
E^\prime=E-\mathbf{v}\cdot \mathbf{P}+\frac{1}{2}Mv^2.
\end{align}
Here primed coordinates are quantities in the crystal (co-moving) frame and unprimed coordinates are in the lab frame. Using results (\ref{Emovingwall}) and (\ref{DEP}) we obtain 
\begin{align}
E^\prime(\mathbf{q})=E_0-\Delta E(\mathbf{q}),
\end{align}
where $\Delta E$ defined in  (\ref{DEP}), and we have changed variable $\mathbf{v}\to\mathbf{q}$. From this it is easily verified that the effective mass definition (\ref{fijmstar}) of the superfluid fraction tensor is identical to Eq.~(\ref{fijE}).
   
 \section{Superfluid fraction results}
In this section we evaluate the superfluid fraction for the various modulated states of the 2D soft-core gas. We perform these calculations by solving for the auxiliary vector function and then using (\ref{fijP}) to determine $f_{ij}$. We find identical results using the effective mass approach. The numerical calculations are preformed in a single unit cell using the spectral approach discussed in Sec.~\ref{Sec:2Dmodel}.

\begin{figure}[htbp]
	\centering
	\includegraphics[width=3.35in]{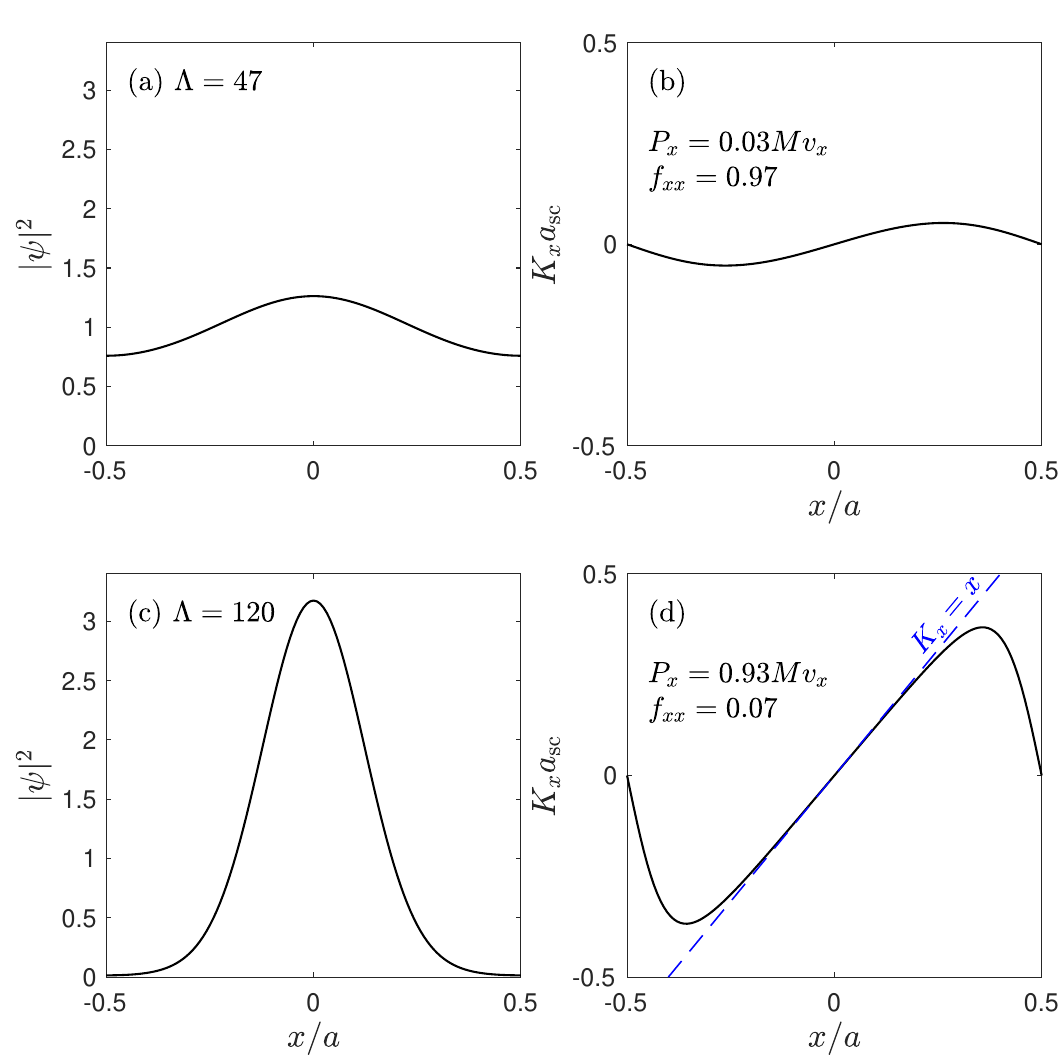}
	\caption{Ground state profiles (a),(c) and $K_x$ auxiliary functions (b), (d) for the stripe state.  Dashed line in (d) shows that $K_x\approx x$ near the center of the cell where the  crystal density is high.
	\label{figstripeSF1}}
\end{figure}

\subsection{Stripe state}\label{sec:stripe}
The stripe state realizes a 1D supersolid, in that crystalline order is along one spatial dimension, here taken to be $x$. For this case the solution of Eq.~(\ref{Ki}) is analytic. Notably,  $K_y$ is zero, $K_x$ reduces to a function of $x$, and  Eq.~(\ref{Ki}) simplifies to
 \begin{align}
\frac{d}{dx} \left(\rho_0\frac{dK_{x}}{dx}\right)=\frac{d\rho_0}{dx}.
 \end{align}
 This has the solution $ K_x(x)=\int_{-\frac{a}{2}}^{x}dx^\prime\left(1-\frac{c}{\rho_{0}(x^\prime)}\right)$, where the integration constant is  $c= [{1}/{a}\int_{-\frac{a}{2}}^{\frac{a}{2}} {dx^\prime}/{\rho_{0}(x^\prime)} ]^{-1}$ to ensure that $K(x)$ is periodic. Using Eq.~(\ref{fijP}) we obtain the tensor components to be
 \begin{align}
 f_{xx}&= \frac{1}{n}\left[\frac{1}{a}\int_{-\frac{a}{2}}^{\frac{a}{2}} \frac{dx^\prime}{\rho_{0}(x^\prime)}\right]^{-1},\label{fsLegg}\\
  f_{yy}&=1,\\
   f_{xy}&=f_{yx}=0,
 \end{align}
 with (\ref{fsLegg}) being the Leggett upper bound for the superfluid fraction \cite{Leggett1970a,Leggett1998a}, which is known to be exact for 1D cases \cite{Sepulveda2008a}.  
 
Examples of the stripe state density profile and the $K_x$ functions are shown in Fig.~\ref{figstripeSF1} for a low contrast state close to the phase transition [subplots (a) and (b)] and a high contrast state deep in the crystalline phase [subplots (c) and (d)]. We also show the momentum for the unit cell in terms of the moving frame velocity component $v_x$. In the high contrast state we observe that $K_x\approx x$ in the center of the unit cell where the wavefunction has high density. This leads to the state having a high momentum (i.e.~majority of the system moving with the walls) and hence a low superfluid fraction along $x$.

\begin{figure*}[htbp]
	\centering
	\includegraphics[width=7in]{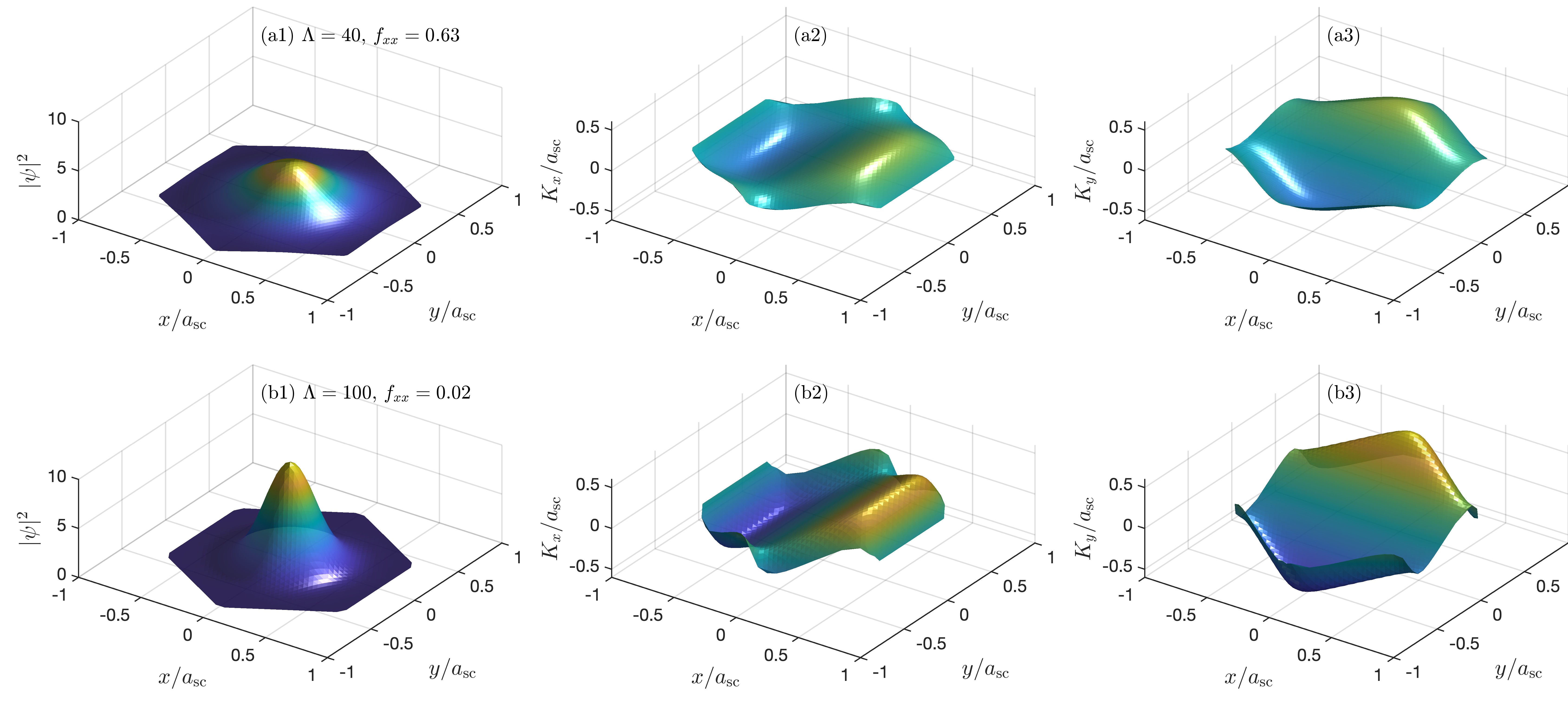}
	\caption{Superfluid response of triangular modulated ground states.  The (a1) ground state density in a Wigner-Seitz unit cell for $\Lambda=40$.  The (a2) $K_x$ and (a3) $K_y$ functions corresponding to the ground state in (a1). The (b1)  ground state  for $\Lambda=100$ and the corresponding (b2) $K_x$ and (b3) $K_y$ functions.  
	\label{figSF1}}
\end{figure*}

\subsection{Triangular and square modulated states}
 We show examples of the triangular crystal ground states and the associated auxiliary vector function in Fig.~\ref{figSF1} for a low contrast case ($\Lambda=40$) close to the transition (a1)-(a3) and a high contrast case ($\Lambda=100$) deep in the crystal regime (b1)-(b3).   
 The $K_x$ function  has an anti-symmetric shape about the cell centre along $x$ [see Figs.~\ref{figSF1}(a2) and (b2)]. The $K_y$ function  is similar to $K_x$, but rotated by 90$^\circ$ (i.e.~anti-symmetric along $y$) [see Figs.~\ref{figSF1}(a3) and (b3)]. For the case with $\Lambda=40$ we have $f_{xx}=f_{yy}=0.63$ and the off-diagonal elements are zero\footnote{Typically we have $f_{xy}<10^{-8}$, which we take as consistent with zero.}.  In all triangular and square lattice cases we find the superfluid fraction tensor to be diagonal and of the isotropic (scalar) form
  \begin{align}
 f_{ij}=f_s\,\delta_{ij}.\label{isotropicSF}
 \end{align}
 This observation of isotropy was also found for the triangular state of the 2D soft-core model in Ref.~\cite{Sepulveda2010a}.  For the case with $\Lambda=100$ the auxiliary function is noticeably sharper [similar to the stripe result in Fig.~\ref{figstripeSF1}(d)] and we observe that $K_x\approx x$ and  $K_y\approx y$ in the central part of the unit cell where the density is high. The superfluid fraction tensor for this case is $f_{s}=0.02$.

 \begin{figure}[htbp]
	\centering
	\includegraphics[width=3.4in]{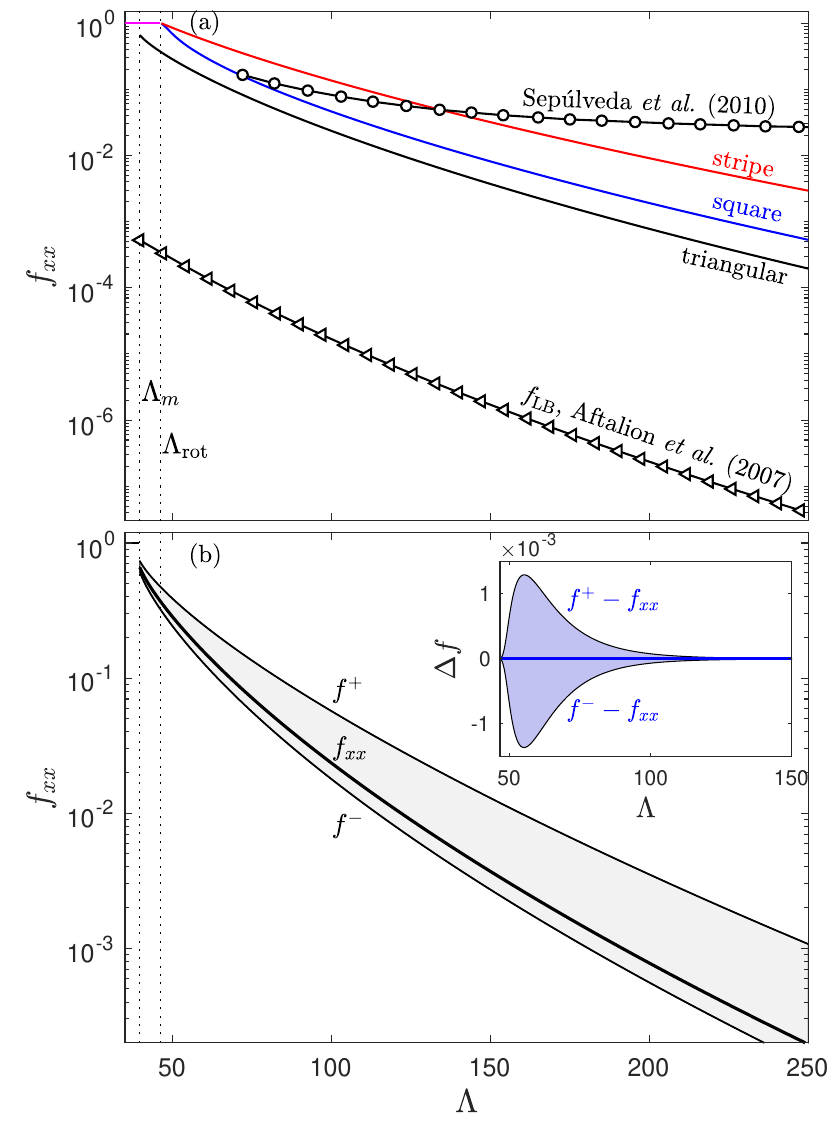}
	\caption{(a) Superfluid fraction $f_{xx}$ computed using (\ref{fijE}) for the various geometry stationary states (solid lines) as a function of $\Lambda$. The data from Sep{\`u}lveda \textit{et al.~}\cite{Sepulveda2010a} for the triangular state presented for comparison (black line squares) and the relevant analytic bound of Aftalion \textit{et al.~}\cite{Aftalion2007a}. (b) The Leggett bounds $f^{\pm}$ shown relative to the  triangular state results from (a). The Leggett bounds are tighter for the square state and are shown in the inset.  
	\label{figSFfrac1}}
\end{figure}
   
\subsection{Superfluid fraction results}
In Fig.~\ref{figSFfrac1}(a) we show the superfluid faction $f_xx$ for a range of $\Lambda$ values and for the three different lattice geometries we consider. The triangular state undergoes a first order transition from the uniform state at $\Lambda=\Lambda_m$, and here the superfluid fraction suddenly drops from unity. In contrast, the stripe and square states undergo a continuous transition at $\Lambda=\Lambda_{\mathrm{rot}}$ and the superfluidity changes continuously. The stripe state has a larger $f_{xx}$ superfluid fraction than the square state, consistent with the stripe state having a lower density contrast relative to the square state [see Fig.~\ref{figCfs}(c)], and hence more tunnelling between unit cells.

For comparison we show the superfluid fraction results (also for $f_{xx}$) reported in Fig.~3 of Sep\'ulveda \textit{et al}.~\cite{Sepulveda2010a}. These results are also for the triangular state and were computed using the auxiliary vector function approach we employ, but are seen to compare poorly to our results. In that work the ground states were calculated on a large square region containing many unit cells, but not commensurate with the unit cells. This could strain the resulting crystal and effect the superfluidity (see discussion in Sec.~\ref{Sec:anisoSF}). Also, as our results show, the auxiliary vector functions exhibit quite sharp features for large $\Lambda$ [e.g.~see Figs.~\ref{figSF1}(b2) and (b3)], which make them difficult to calculate accurately.  In Ref.~\cite{Sepulveda2010a} a finite element methods was used to calculate the auxiliary vector functions, whereas all our calculations are performed using a spectral method (i.e.~planewave expansion in reciprocal lattice vectors) that is specific to the unit cell. We check that our results are converged with respect to the number of reciprocal lattice vectors (typically we use $>10^3$ planewaves in the expansion). Also, we have validated details of our calculations (e.g.~using predictions of the melting point, excitation spectra, optimal lattice parameters) against more recent calculations on the 2D soft-core model \cite{Kunimi2012a,Prestipino2018a}.

In Fig.~\ref{figSFfrac1}(a) we also show the lower bound for the superfluid fraction by Aftalion \textit{et al.~}\cite{Aftalion2007a}. Their work was developed for soft-core models, with the 2D result we show here being
\begin{align}
f_{\mathrm{LB}}=\frac{\Lambda^{2}}{64}e^{-2\sqrt{\frac{\Lambda}{\pi}}\frac{a}{a_{\mathrm{sc}}}}.
\end{align}
Our results in Fig.~\ref{figSFfrac1}(a) confirm that this is a lower bound, but is more than two orders of magnitude smaller than our calculated superfluid fraction for the interaction parameter regime we consider.

\subsection{Anisotropic 2D Superfluidity}\label{Sec:anisoSF}

\begin{figure}[htbp]
	\centering
	\includegraphics[width=3.4in]{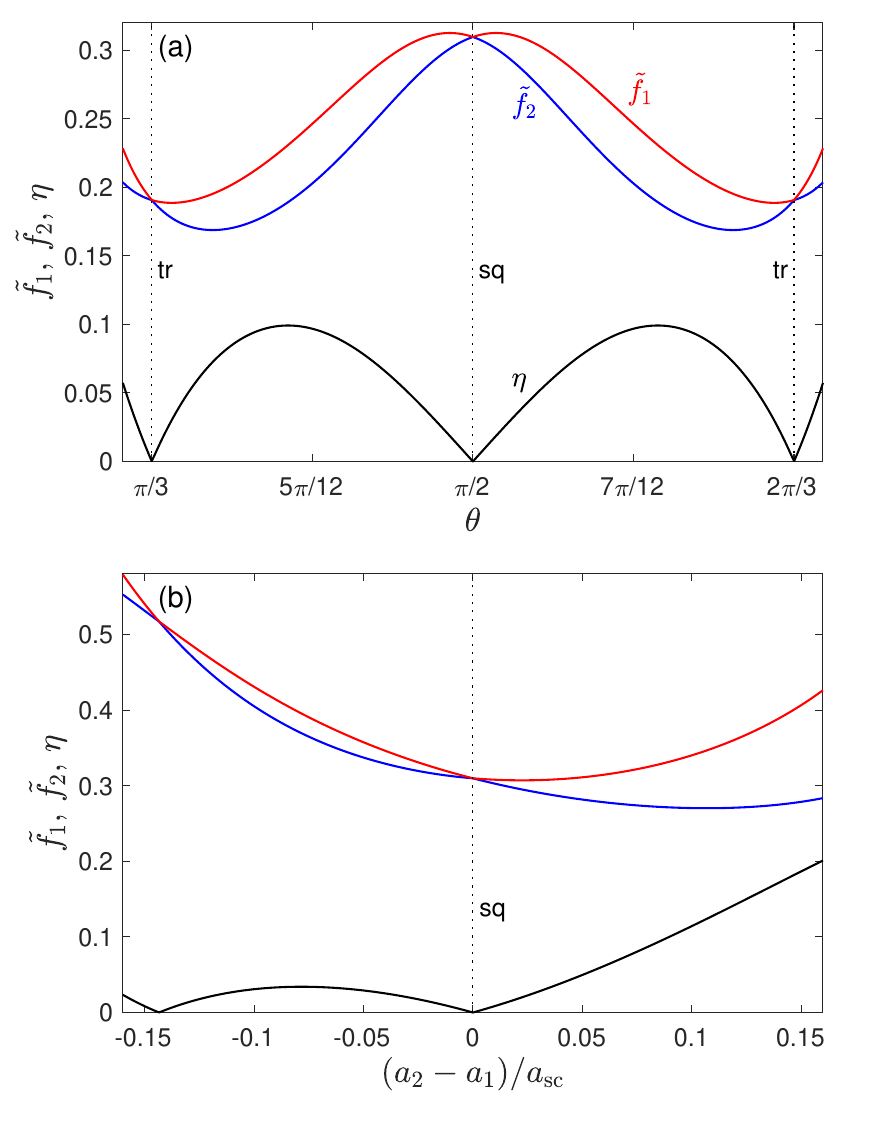}
	\caption{Anisotropic superfluidity characterized by the superfluid fractions $\{\tilde{f}_1,\tilde{f}_2\}$ and anisotropy parameter $\eta=(\tilde{f}_1-\tilde{f}_2)/(\tilde{f}_1+\tilde{f}_2)$. (a) Results for unit cell with varying $\theta$  where $a_1=a_2=1.377\,a_\mathrm{sc}$ is the optimal value for the square lattice. (b) Results for a rectangular cell ($\theta=\pi/2$)  as $a_2$ varies, with $a_1=1.377\,a_\mathrm{sc}$. Results are computed from stationary states  with $\Lambda=60$.
	\label{figanisoSF}}
\end{figure}

Our earlier results showed that for the 2D crystalline square and triangular states states have an isotropic superfluid response (\ref{isotropicSF}), whereas the 1D stripe phase is anisotropic. We investigate here the conditions where 2D crystals become anisotropic. To do this we can extend our consideration to unit cells of the form  $\mathbf{a}_1=a_1 \hat{\mathbf{x}}$, and $\mathbf{a}_2=a_2(\cos\theta \hat{\mathbf{x}}+\sin \theta  \hat{\mathbf{y}})$, i.e.~allowing the direct lattice vectors to be of different length and for  $\theta$ to vary continuously.  
To analyse these cases we consider the superfluid fraction tensor, and here we denote the matrix representing this as $f$. This is a real symmetric matrix and can be taken to a diagonal form by the similarity transformation $\tilde{f}=R_\varphi fR_\varphi^\dagger$, where   \begin{align}
  R_\varphi= \begin{bmatrix} 
        \cos\varphi & -\sin\varphi\\
        \sin\varphi &\cos\varphi  
     \end{bmatrix}
 \end{align} 
 is the rotation matrix with angle $\varphi$. From this we identify the eigenvalues ($\tilde{f}_1,\tilde{f}_2$) of $\tilde{f}$ as the superfluid fractions along the principal axes, which are orthogonal axes rotated at an angle $\varphi$ with respect to $x$ and $y$.
 
In Fig.~\ref{figanisoSF}(a) we consider a crystalline ground state with the lengths of the unit cell fixed ($a_1=a_2$), but allow the angle $\theta$ to vary. The results show that  the isotropic condition ($\tilde{f}_1=\tilde{f}_2$) occurs for  triangular  $(\theta=\pi/3,2\pi/3)$ and square  $(\theta=\pi/2$) lattices, but is anisotropic at other angles. For these results  the rotation angle defining the principal axes is $\varphi=\theta/2$.

 In Fig.~\ref{figanisoSF}(b) we fix the angle to the rectangular case $(\theta=\pi/2$)  and vary the length of $a_2$ relative to $a_1$. In general a difference in length causes the superfluid tensor to become anisotropic, although in this case the principal axes remain aligned to $x$ and $y$ (i.e.~$\varphi=0$). In addition to the square case ($a_2=a_1$), we also observe that an isotropic response occurs at $a_2-a_1\approx-0.144a_{\mathrm{sc}}$.

\section{Leggett bounds for an isotropic 2D crystal}
 \begin{figure}[htbp]
	\centering
	\includegraphics[width=3.4in]{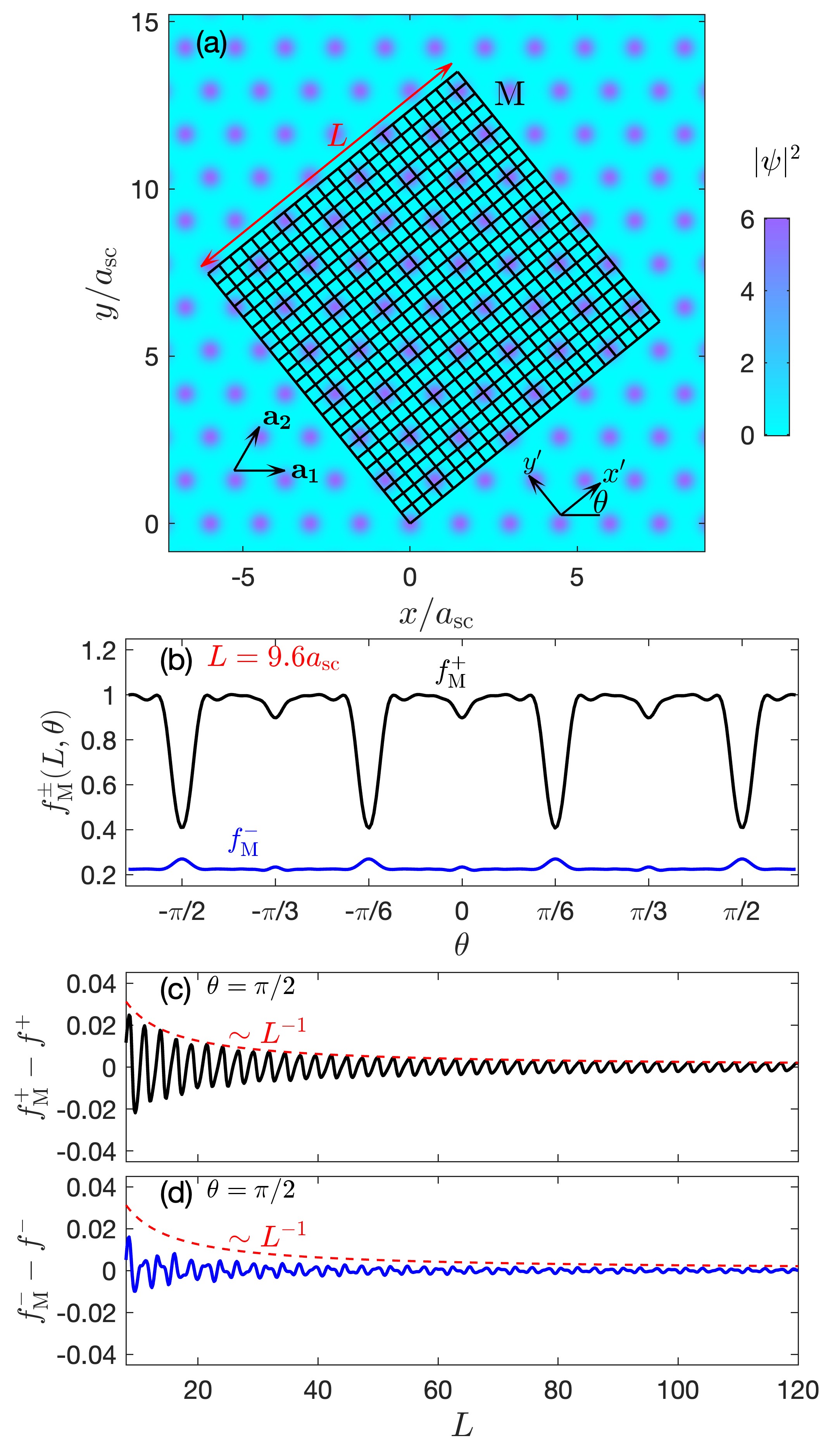}
	\caption{Leggett  bounds. (a) Our macroscopic sample M is specified by square grid of dimensions $L\times L$ indicated by the mesh. We have introduced the coordinates $(x^\prime,y^\prime)$, which are rotated by $\theta$ relative to the default axes. The crystalline density over this region is integrated along $x^\prime$ and $y^\prime$ to evaluate the Leggett bounds $f_{\mathrm{M}}^{\pm}$. (b) The Leggett upper  (black line)  and lower (blue line) bounds for $L=9.6a_{\mathbf{sc}}$ as a function of $\theta$. The local minima of  $f_{\mathrm{M}}^{+}$  and local maxima of  $f_{\mathrm{M}}^{-}$ occur when the rotated axes are parallel to the  direct lattice vectors. The  (c) $f_{\mathrm{M}}^+$ minima (least upper bound) and (d) $f_{\mathrm{M}}^-$ maxima (largest lower bound) at $\theta=\pi/2$ as a function of $L$, relative to the unit cell values [from Sec.~\ref{Sec:Legguc}]  $f^+=0.4283$ and $f^-=0.2794$, respectively. The triangular ground state for $\Lambda=48$ is used for the results presented in this figure.
	\label{figLeggett1}}
\end{figure}
 
Leggett has developed bounds for the superfluid fraction based on integrals of the ground state density \cite{Leggett1970a,Leggett1998a} (also see  \cite{Eggington1977a}).  
  Here we specialize these bounds to the 2D case and restrict our attention to cases of isotropic superfluidity (i.e.~triangular and square crystals). We first investigate them in application to a \textit{macroscopic} region (with many unit cells), in the spirit of Leggett's original formulation. We use this to investigate the sensitivity to size and orientation of the region. These results motivate us to specialize the bounds to a form that can be evaluated, without any adjustable parameters, using the system density on a single unit cell.
  
 \subsection{Application of a macroscopic region}
 Here we consider evaluating the Leggett bounds using the system density profile in a square macroscopic region M of size $L\times L$, with $L\gg a$ (i.e~enclosing many unit cells). Such a region is indicated in  Fig.~\ref{figLeggett1}(a)\footnote{In practice a much denser grid of points is used to sample the density to ensure discretization errors are negligible.}. The expressions \begin{align}
 f_\mathrm{M}^+(L,\theta)&\equiv\frac{1}{n}\left[\frac{1}{L}\int^L_0\frac{dx^\prime}{\frac{1}{L}\int^L_0dy^\prime|\Psi(\mathbf{x}^\prime)|^2}\right]^{-1},\\
 f_\mathrm{M}^-(L,\theta)&\equiv\frac{1}{n}\frac{1}{L}\int^L_0dx^\prime\left[\frac{1}{L}\int^L_0\frac{dy^\prime}{|\Psi(\mathbf{x}^\prime)|^2}\right]^{-1},
 \end{align}
are the upper and lower  bounds, respectively.
 We vary the orientation of the region M by rotating the square region by an angle of $\theta$ with respect to the standard axes [see Fig.~\ref{figLeggett1}(a)]. As the angle changes relative to the crystal axes (defined by $\mathbf{a}_1$ and $\mathbf{a}_2$) the value of $f_{\mathrm{M}}^{\pm}$ changes. We notice that there are minima (maxima) in $f_{\mathrm{M}}^{+}$ ($f_{\mathrm{M}}^{-}$) when the axis of the first integration  (i.e.~$y^\prime$-axis) is parallel to  the direct lattice vectors  [e.g.~it is parallel or anti-parallel to $\mathbf{a}_1$ when $\theta=\pm\frac{\pi}{2}$, and similarly for $\mathbf{a}_2$ when $\theta=\pm\frac{\pi}{6}$, etc., see Fig.~\ref{figLeggett1}(b)]\footnote{We note that similar results hold for the square lattice case, except that the minima of $f_{\mathrm{M}}^{\pm}$  occur at $\theta=0,\pm\frac{\pi}{2},\ldots$.}. Thus, we can find the strictest bounds by selecting $\theta$ to align to these directions. In Figs.~\ref{figLeggett1}(c) and (d) we consider how $f_{\mathrm{M}}^{\pm}$, evaluated at  $\theta=\pi/2$, changes with increasing size of the integration region.   The results oscillate, but converge as $L$ increases. These oscillations occur because the integration region is not commensurate with the unit cells, and suggest that we would be able to extract the value from a careful analysis of a single unit cell.

 \subsection{Application to a unit cell}   \label{Sec:Legguc}
 \begin{figure}[htbp]
	\centering
	\includegraphics[width=3.4in]{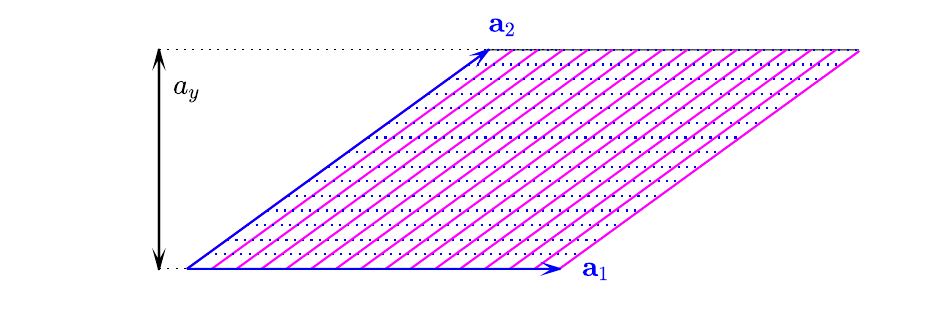}
	\caption{Primitive unit cell geometry relevant to defining $f^\pm$. 
	\label{figaffine1}}
\end{figure}
For the 2D cases we  considered in this paper,  the  primitive unit cell defined by $\mathbf{a}_1$ and $\mathbf{a}_2$ is a rhombus with side length $a$, angle $\theta$ and vertical height $a_y=a\sin\theta$. We specialize the Leggett bound definitions to the unit cell as follows
 \begin{align}
 f^+&\equiv \frac{1}{n} \left[\,\,\int\displaylimits_{\mathbf{a}_1}\frac{dx}{a}\frac{1}{\int\displaylimits_{\mathbf{a}_2}\frac{dy}{a_y}|\Psi(\mathbf{x})|^2}\right]^{-1},\label{f+}\\
 f^-&\equiv \frac{1}{n}\int\displaylimits_{\mathbf{a}_1}\frac{dx}{a}\left[\,\, \int\displaylimits_{\mathbf{a}_2}\frac{dy}{a_y}\frac{1}{|\Psi(\mathbf{x})|^2}\right]^{-1}.\label{f-}
 \end{align}
Here the integral $\int\displaylimits_{\mathbf{a}_j}$ refers to integration parallel to the direct lattice vector $\mathbf{a}_j$, i.e.~along the horizontal dotted blue lines or diagonal magenta lines in Fig.~\ref{figaffine1} for $\mathbf{a}_1$ and $\mathbf{a}_2$, respectively. This turns to be be convenient numerically, as the Fourier based sampling of the wavefunction is on a grid of the type represented in that figure. Hence these integrals can be numerically calculated with spectral accuracy using modest resources. As verified in Figs.~\ref{figLeggett1}(c) and (d), the macroscopic application of the Leggett bounds at the optimal angles converge to $f^\pm$ as $L\to\infty$.

To make the  integrals appearing in  Eqs.~(\ref{f+}) and (\ref{f-}) unambiguous we note that the unit cell can be mapped to a rectangular region of dimensions $a\times a_y$ with the affine transformation  (a shear in the $x$-direction) of
  \begin{align}
  A_t= \begin{bmatrix} 
        1 & -\cot\theta\\
        0 &1 \\
     \end{bmatrix}
     \end{align} 
Applying the affine transform we define the  coordinates $\mathbf{u}=A_t\mathbf{x}$, with $\mathbf{u}=(u_x,u_y)$. The primitive unit cell becomes  the  rectangular region $-\frac{1}{2}a\le u_x\le\frac{1}{2}a$ and $-\frac{1}{2}a_y\le u_y\le\frac{1}{2}a_y$. Noting that the Jacobian determinant of this transformation is unity, the integrals can be written as, for example
   \begin{align}
   \int\displaylimits_{\mathbf{a}_2}dy\,|\Psi(\mathbf{x})|^2\equiv\int^{+\frac{a_y}{2}}_{-\frac{a_y}{2}}|\Psi(A_t^{-1}\mathbf{u})|^2du_y.
   \end{align}

\subsection{Comparison to exact calculations}   
In Fig.~\ref{figSFfrac1}(b) we compare $f^\pm$ to our earlier results for $f_{xx}$ superfluidity for the triangular case. For larger $\Lambda$ values $f^+$ can be an order of magnitude larger than $f^-$, so the bounds are not that tight, but still provide a useful estimate. In contrast for the square lattice states [inset to Fig.~\ref{figSFfrac1}(b)] the bounds are much tighter.  Here our results reveal that the bounds estimate the superfluid fraction with an absolute error $|\Delta f|=|f^\pm-f_{xx}|\lesssim10^{-3}$, similar to results for a 1D dipolar supersolid in a tube found in Ref.~\cite{Smith2023a}. The bounds are more effective in this case because the wavefunction is close to being separable. For the stripe phase the well-known 1D version of the bound is exact (also see Sec.~\ref{sec:stripe}).

\section{Conclusions}
 
 We have discussed several computational approaches for determining the superfluid fraction tensor of the meanfield system. 
 One approach involves determining the phase response to moving walls by solving two Poisson-like equations for the vector auxiliary equations (\ref{Ki}). From these solutions the superfluid density can be evaluated using Eq.~(\ref{fijE}). The effective mass approach instead involves solving the linear Schr\"odinger problem (\ref{EGPq}), but requires a finite difference in $\mathbf{q}$ to calculate the superfluid tensor. We have discussed why these approaches are formally equivalent. Our results show that the ground state triangular supersolid and the metastable square supersolid both exhibit isotropic superfluidity.
We have investigated how changes in the lattice vector orientation and length can cause the superfluity to become anisotropic.  We also considered how the Leggett bounds can be applied to 2D supersolids with isotropic superfluidity, finding that the tightest bounds from the thermodynamic-limit of their application can be deduced from a calculation on a single unit cell.
 
 Despite the theoretical definitions of superfluidity being well-known for many decades, there are few explicit results for higher dimensional supersolids. Indeed our results disagree with previously published results for the 2D soft-core model, which forms one of the most basic models for describing higher dimensional supersolidity. This lack of accurate numerical results has likely hindered a better understanding of the superfluid response in this system, and how to apply the Leggett bounds in 2D and 3D crystals. In this absence of understanding many proxy-measures of superfluidity have been used in the recent literature without justification (e.g.~various forms of density contrast etc.)

 While our work here has focused on the thermodynamic limit, our results will have implications for finite trapped systems of the type realized in experiments. The  auxiliary vector function calculation only requires a density profile within a unit cell, which could allow this approach to estimate the local superfluid fraction (associated with that cell) in a finite system. Similarly the Leggett bounds could also be applied to this case. Since recent experiments have been able to apply the Leggett upper bound to BECs in 1D optical lattices \cite{Tao2023a,Chauveau2023a}, we expect this avenue of investigation may be practical for experiments considering superfluidity of higher dimensional supersolids (or to BECs inhigher dimensional optical lattices).
 It would be interesting to compare the distribution of local superfluid behavior to global measurements of the trapped case, e.g.~via the calculation of the rotational inertia (e.g.~see \cite{Cooper2010a,Hsueh2012a,Roccuzzo2020a,Tengstrand2021a,Tanzi2021a,Roccuzzo2022a,Gallemi2022a,Norcia2022a}).
 A future direction would be to consider the extension of these ideas to consider superfluidity in the stripe phase of  spin-orbit coupled BECs (c.f.~\cite{Li2013a,Zhang2016a,Chen2018a,Martone2021a}), in which Galilean invariance is also broken. 
It might also be necessary to explore scenarios in which the supersolid state may not necessarily be the ground state, but rather exhibits specific phase excitations \cite{Petter2021a} (also see \cite{Zhang2020a}) or lattice distortions. In the latter case, our results already suggest this will lead to anisotropic effects in the superfluid response.

\section*{Acknowledgments}
 The author would like to acknowledge communications with Amandine Aftalion, Mayasa Kumini and Yong-Chang Zhang about their work, and useful discussions with Danny Baillie, Lauriane Chomaz, Francesca Ferlaino, Au-Chen Lee, Sukla Pal, Lily Platt, and Ben Ripley.
%

\end{document}